\documentstyle[prb,aps,twocolumn,amssymb,amsfonts,epsfig]{revtex}
\begin{document}

%----------------------------------------------------------------------
% Title

\title{Electronic structure and magnetism in X$_x$W$_{1-x}$O$_3$ (X=Nb,V,Re) from supercell calculations.}

\author
{T. Jarlborg}

\address{D\'epartement de Physique de la Mati\`ere Condens\'ee,
Universit\'e de Gen\`eve, 24 Quai Ernest Ansermet, CH-1211 Gen\`eve 4,
Switzerland} 

\date{\today}
\maketitle

%----------------------------------------------------------------------
% Abstract
 Some doped semiconductors have recently been shown to
 display superconductivity or weak ferromagnetism.  Here
 we investigate the electronic structure and conditions for magnetism in a supercells of 
 cubic XW$_{26}$O$_{81}$,
 where X=Nb,V and Re.
 The undoped material is an insulator, and although the slightly doped material is a metal, it is far
 from the Stoner criterion of magnetism. The conditions of a localized density-of-states (DOS) which
varies rapidly with the energy, resemble those of doped hexaborides. The virtual crystal approximation
is used to vary the doping level. A small moment appears if
the Fermi energy, $E_F$, coincides with a large derivative of the DOS.  
\begin{abstract}

\end{abstract}

\pacs{75.50.Cc,71.15.Mb}

%----------------------------------------
\section{Introduction.}

 Superconductivity and magnetism are expected to compete in materials having a large density-of-states
 (DOS) at $E_F$. However, recent studies of low DOS materials have
 found ferromagnetism (FM) of an unusual form, where a weak moment is combined with a surprisingly high
 Curie temperature, $T_C$. There are also examples of superconducting systems with a low DOS
 at $E_F$, since they are doped insulators \cite{raub,salje,reich}.  
 In particular, the quest of finding systems similar to the electron doped WO$_3$,
 where small additions of Na are reported to make the system superconducting \cite{salje,reich}, 
 suggests that hole doping should work as well. 
 Instead, doping by dilute (2.5 \%) Nb substitutions on W sites leads to weak itinerant ferromagnetism \cite{fel},
 with properties not too different from those of doped hexaboride systems \cite{you}. In this work
 we study the electronic structure of cubic XW$_{26}$O$_{81}$, with X=Nb, V and Re, 
 in order to see if similar conditions for magnetism
 are met as in La doped strontium hexaboride \cite{hexlet}.
 
 The search for weak magnetism in metal doped oxide insulators is motivated by the hope
 that the correction of the 
Stoner factor by the Coulomb term \cite{hexlet} should push some systems towards a FM state. The
Stoner criterion for magnetism reads as follows when it is corrected for Coulomb energies;
\begin{equation}
\label{eq:S}
\bar{S}_U = N I_s + U_0 N' N_v / N^2
\end{equation} 
where $I_s$ is the usual exchange integral \cite{jan,c15},
$N$ is the density of states at E$_F$, $N'$ is the energy derivative of the DOS of
the impurity band, and $N_v$
is the DOS of the remaining bands.  The parameter $U_0$
is the Coulomb energy associated with a transfer of $\Delta q$ electrons to the impurity. 
Magnetism is possible when $\bar{S}_U>$1.
It is only when the impurity band  has a large energy derivative of the DOS,
that charge transfers (CT) can be activated by an imposed exchange splitting.  The last term of
eq. 1 is negligible in most system, so that $\bar{S}_U$ becomes the usual Stoner factor. But this term can be
important because of various reasons. It depends on
the sign of $U_0$ if the Stoner factor will increase or decrease because of the CT mechanism. It is difficult
to calculate
$U_0$ from a model DOS, because its value depends on self-consistent
screening. But if the tendency follows eq. 1 one expects a large effect either because of large $U_0$, $N'$ and
$N_v$, or because of low $N^2$. If the background DOS is dominant we have $N_v \approx N$, so the condition
is that $U_0 N' / N$ should be large. If an impurity band is formed at the band edge either below or
above the gap in an insulator, one can expect a large ratio of $N'/N$ because the total DOS will
be small for small concentrations of impurities, while the slope of the DOS of
the impurity band can be relatively high. The amplitude and sign of $U_0$  
can depend on the dopant as well as on the host material.
We will do self-consistent, spin-polarized calculations both for hole and electron doping. If the CT are
opposite for the two types of doping and if $U_0$ is of the same sign for several types of dopants, one
should find different behavior for the magnetic ordering.

\section{Method of calculation.}

The electronic structures are calculated
using the self-consistent linear muffin-tin orbital method in the local spin density approximation
\cite{lmto,lda}. 
  The calculations are made for the cubic, perovskite type structure of WO$_3$, shown in figure 1. 
The oxygens occupy the face centered positions
and the tungsten atom is at the center. No atoms occupy the open space at the cube corners, but
the calculations include empty spheres at these positions. 
 The supercell includes
27 elementary cells (or formula units, f.u.), 
where the 4-atom cells have been tripled along x, y and z. The exchange of one W with Nb is done
in the central cell, making the concentration $x$ close to 0.04. 
The calculations involve totally 135 sites (108 atoms and 27
empty spheres) of which there are 14 nonequivalent ones. The basis-set includes s,p, and d for all sites, 
with f included in tails,
and the bands are determined in 4 or 10 irreducible k-points. The virtual crystal approximation (VCA) is used in
two cases with Nb as dopant. The nuclear charges and the number of electrons are increased to a non-integer
number. As discussed later, each atom will receive 0.002 or 0.005 additional charges in VCA calculations
with Nb as dopant.
Other details of the calculations are the same as in the work on electron doping of WO$_3$ and SrTiO$_3$ \cite{stolet}.

The real structure of Nb$_x$W$_{1-x}$O$_3$ in the experiment of Felner et. al.  \cite{fel} is tetragonal,
where the concentration of Nb is lower ($x \approx 0.025$) than in the supercell here, and it 
may have some oxygen deficiency. The goal
of the present work is not to study exactly the system used in the experiment, but
to find out whether the conditions for weak ferromagnetism can be found in doped oxides
as was found in some hexaborides \cite{hexlet},
 and if such conditions can be expected to be frequent around impurities in metallic compounds.
As will be shown, the result depend on details of the impurity DOS near E$_F$, 
so that magnetism appears to be correlated
with a large derivative ("slope") of the DOS of the impurity band, as was found in the case of hexaborides. 
The conditions for this
to happen in real materials are delicate.  Large slopes of the DOS of the impurity band are needed and this can
be a limiting case for very low impurity concentrations,
 when the size of the supercell and disorder of the real
material will prohibitive for computations.  

\section{Results.}

 The calculated DOS for NbW$_{26}$O$_{81}$ and undoped WO$_3$ (W$_{27}$O$_{81}$) 
 near the valence band edge are shown in fig. 2. 
 The band gap of about 2.1 eV for pure WO$_3$ is reduced to about 1.5 eV
because of a separated NbO band complex just above the rest of the WO valence bands, and
  $E_F$ enters into the valence band below the gap.  A close up of the total DOS of NbW$_{26}$O$_{81}$
  is shown in fig. 3, together with the partial DOS functions from the Nb impurity, from the six closest O,
   and from all W. The partial
 DOS functions show that the state just below the gap is very different from the uniform W-O 
 mixture in the undoped system.
The Nb impurity acts very differently from a W atom, and makes the p states on the nearest oxygens surrounding the 
impurity to become localized.
It is seen from fig. 3 that about one third of the total DOS at $E_F$ is coming from the six O-neighbors.
The local DOS on the Nb impurity is only about 3 percent of the total DOS, cf. Table 1.

 These results are very different from those of
electron doping with Nb substitution on a Ti site in SrTiO$_3$ \cite{stolet}. The local DOS on Nb and 
the surrounding oxygens in that case are very close to the respective
DOS on Ti and O far away from the impurity. In other words, a 
rigid-band model applies to Nb doping in SrTiO$_3$, but not to Nb in WO$_3$.

The moment tends to zero (or below 0.02 $\mu_B$/cell) when one W is replaced with Nb. According to the mechanism
of CT there should be a correlation between charge transfer, magnetic moment and the slope of the
partial DOS at E$_F$. As seen in fig. 3, E$_F$ is at a flat region of the DOS.
In order to bring E$_F$ into the negative slope of the DOS closer to the band edge,
we use the VCA in two sets of calculations in order to add 0.002 and 0.005 electronic charges per atom.
This adds 0.216 and 0.54 electrons per cell, respectively. This is a small amount of additional charge, and the 
paramagnetic DOS is almost identical to the one shown in fig. 3, except that E$_F$ is moved
a few mRy closer to the gap. The self-consistently calculated moment in the spin-polarized
calculations tend to finite values in these cases. When 4 k-points are used we find about 0.72 $\mu_B$/cell
when 0.216 electrons are added. This moment decreases when the calculations are continued using 10 k-points,
but the moment becomes stable near 0.20 $\mu_B$/cell, as seen in Table 1. 
These results confirm the mechanism of having a stabilizing effect from charge transfer when the band
structure becomes spin polarized.

One calculation is made using 4 k-points in the case
of 0.54 additional electrons, where E$_F$ is moved closer to the gap.  The value of $N'$ is about twice as large
at 0.54 as at 0.216 additional electrons, but the total DOS is lower, 88 instead of 102 states/Ry/cell.
The spin-polarized calculation converges to a moment of about 0.28 $\mu_B$/cell. 
This example seems to contradict the assumed correlation between $N'$ and m (because the moment is lower than 
with 0.216 additional electrons), but it is clear that the relation
does not hold for a vanishing DOS very close a gap, where nonlinear variations of the DOS becomes important.

Calculations for an impurity of V instead of a Nb give results in the DOS which are
quite similar. But there are changes in details, which have an 
influence on the results for the magnetic moment. The smaller V atom makes the band edge different so that
the DOS has a negative slope at E$_F$ (cf. fig. 4 and Table 1) 
instead of being almost constant in the case of a Nb impurity. The DOS of the six oxygen atoms close to the V site 
is clearly negative and an increase of temperature or an imposed exchange splitting should
decrease the charge on these atoms. Spin-polarized calculations give
a considerable moment, about 0.92 $\mu_B$/cell, when they are converged with 4 k-points.
Note that this is without rigid-band shifts of E$_F$ via VCA calculations.
In the case of Nb it was not possible to find a stable moment unless a change in doping was forced by
addition of 0.2 to 0.5 electrons per cell via the VCA.

Finally, a calculation is made for ReW$_{26}$O$_{81}$.  It could be expected that the additional electron
in Re (compared to W) will be added to the conduction band in a rigid band manner
as for Nb in SrTiO$_3$ \cite{stolet}. If so it could lead to an opposite condition for the correction
to the Stoner factor, since $N'$ should be positive. However, the results
from the band calculations show a very different situation.
The Re-d band becomes localized just at the top of the valence band and
hybridizes with O-p, as shown in fig. 5, instead of appearing at the bottom of the
conduction band. The
DOS near E$_F$ is large on the impurity site (mainly Re-d). This is different to the cases of Nb and V as
impurities, where the DOS on the impurity site is very small, while the six nearest oxygen sites take a significant
part of the DOS. Other differences with the cases of Nb and V is that the derivatives
of the total and partial DOS are varying rapidly with energy, and the amplitude of the total DOS is larger. 
From fig. 5 it is seen that $N'$ near E$_F$ is generally positive, even though a local peak just above
E$_F$ makes the conditions
for charge transfer very complex. An increase of the temperature or an exchange splitting should mainly
lead to a CT going in the opposite direction from what is the case with Nb or V impurity, although non-linearities
are probable. This difference with Nb and V suggests that the magnetic state should not appear with Re as
an impurity. Spin-polarized calculations using 4 k-points,
 where the self-consistent iterations start from an imposed large moment, 
converge towards a vanishing moment, smaller than 0.025 $\mu_B$/cell. It can be concluded that
no stable moment exists in this case.

\section{Discussion.}

These calculations show that there is a possibility of weak magnetic states around dopants of Nb and V in
WO$_3$, and as in the case of La dopant in hexaborides,  there is a correlation
between moment and sharp DOS structures of the impurity band. However, a notable difference with
the hexaboride is that the impurities provoke a large localized DOS not on the impurity itself,
but on the 
nearest oxygen sites, so that the O$_6$ band plays the role of impurity band.
  When there is a moment it is located ($\sim 80$ percent of the total moment) to the O$_6$ cluster
  surrounding the impurity.
These results suggest that the appearance of weak local moments can be quite common in many
doped insulators. However, weak ferromagnetism are found at very low doping concentration
in real, structurally distorted systems, and the present
calculations are not exactly applicable for such systems. The present type of calculations demonstrate
a mechanism behind the appearance of local moments, but whether the conditions of sharp DOS structures
are sufficient for magnetism in each individual case is a delicate question. For instance, the previous study of
La doped SrB$_6$ found a stable moment for a cell containing 8 formula units, while with one impurity in a
27 f.u. cell did not. 
A calculation with a rigid band shift of E$_F$ within the VCA could restore favorable conditions 
for a small moment \cite{hexlet}.
 The Curie temperatures associated with
the small moments are not yet understood within the 
present mechanism \cite{mcomm}.

Semiconductors can be made magnetic by doping of magnetic ions, like Mn in GaAs, where the magnetic RKKY
interaction is believed to responsible for the weak polarization of the host \cite{ohno}. Band calculations
using LDA for Mn impurities find a magnetic moment, which is localized on the Mn site \cite{mac}.
This is somewhat different from the situation here, where a nonmagnetic ion make an insulating material
weakly magnetic. The RKKY interaction might be essential for the polarization of the host, but the
question is why the impurity or the closest neighbors become magnetic when the none of the ingredients
(impurity and host) appear magnetic separately. This will be a Stoner-like case, where the DOS
is such that the total energy is lower for the magnetic configuration, and this is irrelevant of
having to deal with magnetic or nonmagnetic ions. It is through the DOS and the total energy
of the compound that the magnetic state can be understood. A good example of this can be found in the calculated
structural variation of magnetism in iron. The common bcc phase is magnetic, while in the fcc or
hcp phases, at ambient or high pressure, the ground state is nonmagnetic. It is therefore not evident
that Fe should be a magnetic in all host materials, like it is not evident that Nb as an impurity
should cause magnetism in some cases. 
 
In conclusion, from these results we propose that weak FM near impurities can
be caused by a combination of exchange energy and charge transfer energy. The latter energy is favorable
to spin splitting only under some precise conditions of the electronic structure near the impurity. This
makes the appearance of such 'enhanced' Stoner magnetism rather circumstantial, and from the examples studied so far
it can only produce small moments. The T-dependence of the moments is not expected to be anomalous and 
can not explain large Curie temperatures.

%----------------------------------------
% Acknowledgments

%----------------------------------------------------------------------
% Biblio

%----------------------------------------------------------------------
% Table

\begin{table}
\caption{Dopant (X), valence charge (Q), total (N) and partial (on X) and O$_6$, derivative of the total DOS (N'),
 and moments (m). 
The moments are calculated selfconsistently using 10 k-points for Nb and 4 for V and Re.
There is no smooth variation of the DOS near E$_F$ in the case of Re, and N' is not well defined.
Results from VCA calculations are shown in the second line for one case with Nb doping, see the
text.
}
\begin{tabular}{ccccccc}
X & Q & N & N$_{X}$ & N$_{O_6}$ & N' & m  \\
 &  el./cell &  & (cell Ry)$^{-1}$ & & (cell Ry$^2)^{-1}$ & $\mu_B$/cell     \\
\tableline
Nb & 647.0 & 106  & 3 & 36 & -800 & $\sim 0$ \\
Nb & 647.2 &  102   & 3 & 35 & -2900 & 0.20 \\
%Nb & 647.5 &  88   & 2 & 30 & -5900 & 0.28 \\
V  & 647.0 & 145   & 3 & 88 & -8200 & 0.98 \\
Re & 649.0 & 500 & 180 & 92 &  $ >$0   & $\sim 0$ \\
\end{tabular}
\end{table}

%----------------------------------------------------------------------
% figure
\begin{figure}[tb!]
%\begin{center}
\leavevmode\begin{center}\epsfxsize8.6cm\epsfbox{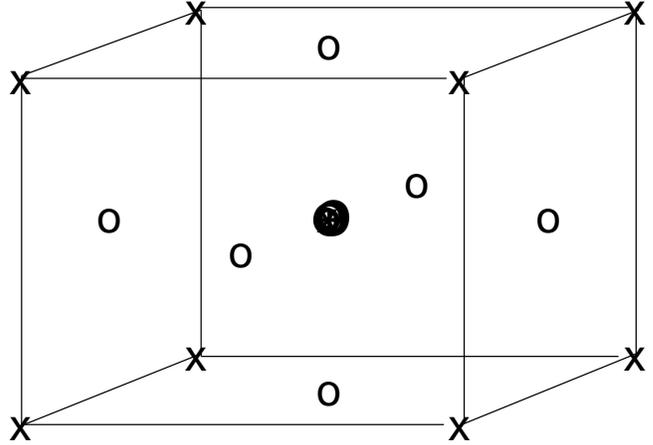}\end{center}
\caption{The basic unitcell of cubic WO$_3$. The oxygens are at the empty circles,
the W at the filled circle, and empty spheres are included in the calculations
at the positions of the crosses. The supercell in the calculations consists of 27 of
the basic unit cells, with one of the W atoms replaced with an impurity site.}
\end{figure}
\begin{figure}[tb!]

% figures directly (hr bw)\begin{figure}[tb!]
\leavevmode\begin{center}\epsfxsize8.6cm\epsfbox{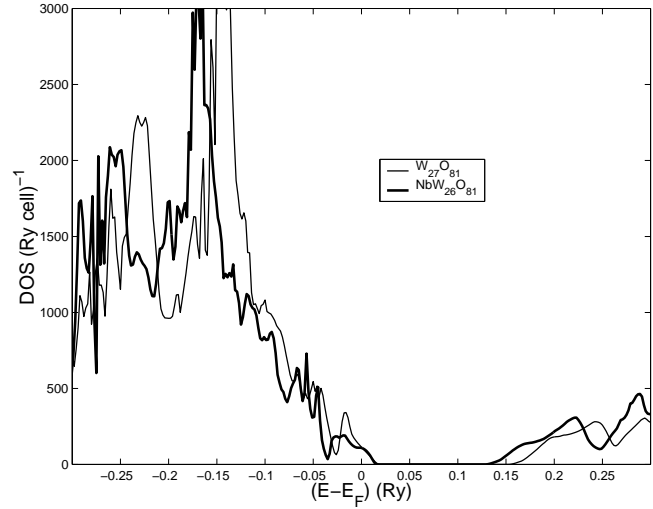}\end{center}
\caption{
 Calculated DOS for W$_{27}$O$_{81}$ and NbW$_{26}$O$_{81}$ using 10 k-points in  the
 irreducible Brillouin zone. The energies are aligned so that zero energy is at a band filling
 of 647 valence electrons in both cases. The Fermi energy is in the middle of the gap for the undoped and at
 zero energy for the doped case.
 }
\end{figure}
\begin{figure}[tb!]
%\begin{center}

% figures directly (hr bw)\begin{figure}[tb!]
\leavevmode\begin{center}\epsfxsize8.6cm\epsfbox{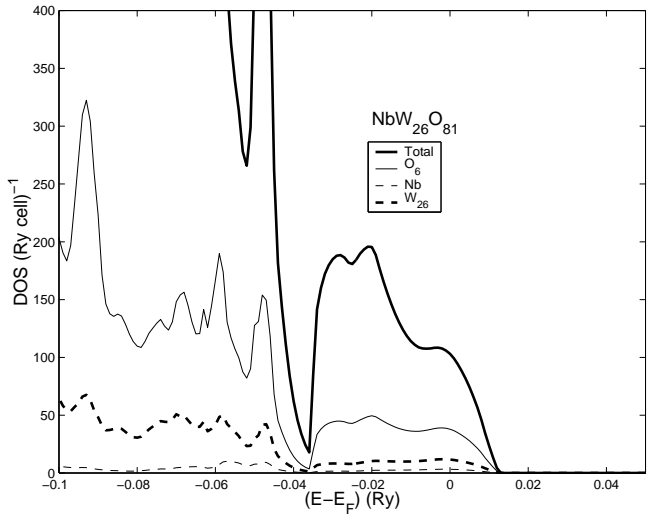}\end{center}
\caption{
 Calculated DOS near E$_F$ for NbW$_{26}$O$_{81}$ and the partial DOS from Nb, W$_{26}$ and the six
 oxygen atoms closest to the Nb impurity. The DOS contains 1 electron between the position of E$_F$
 and the top of the valence band edge. Calculations in the virtual crystal approximation with
 0.216 and 0.54 additional electrons put E$_F$ more to the right, where the negative slope
 is large.
 }
\end{figure}

\begin{figure}[tb!]
\leavevmode\begin{center}\epsfxsize8.6cm\epsfbox{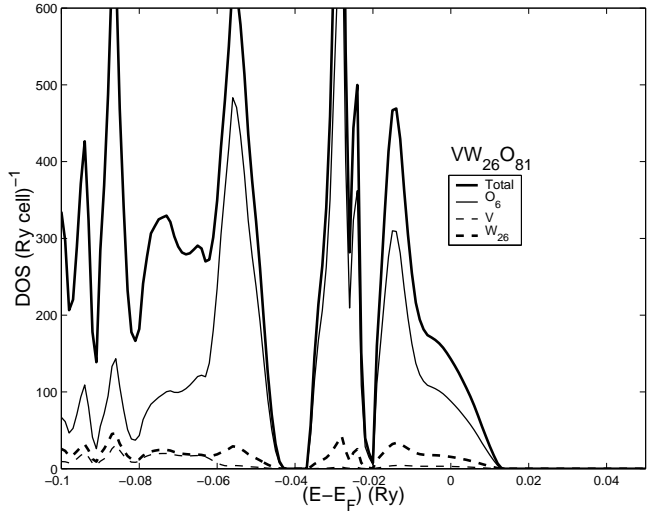}\end{center}
\caption{
 Calculated DOS near E$_F$ for VW$_{26}$O$_{81}$ and the partial DOS from V, W$_{26}$ and the six
 oxygen atoms closest to the V impurity.
 }
\end{figure}

\begin{figure}[tb!]
\leavevmode\begin{center}\epsfxsize8.6cm\epsfbox{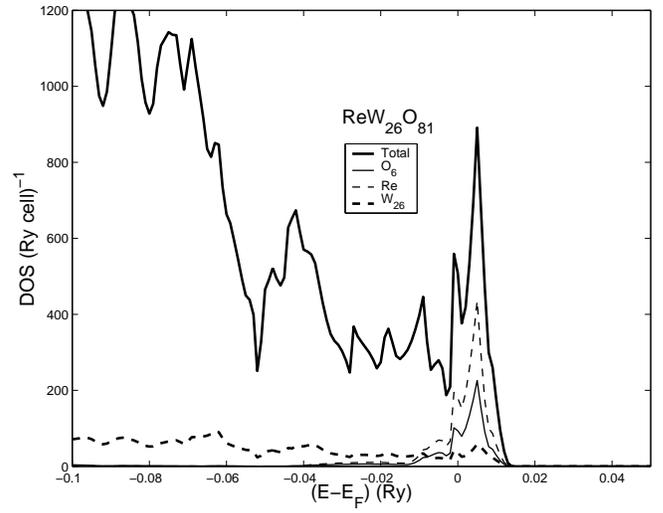}\end{center}
\caption{
 Calculated DOS near E$_F$ for ReW$_{26}$O$_{81}$ and the partial DOS from Re, W$_{26}$ and the six
 oxygen atoms closest to the Re impurity. Note that the Re-d band forms an impurity band at the valence 
 band egde, and thus the rigid band model does not apply in this case.
 }
\end{figure}

\end{document}